\documentclass[10pt, journal]{IEEEtran}

\usepackage{amsthm,amsmath}
\usepackage{longtable}
\usepackage{latexsym}
\usepackage{amscd}
\usepackage{amsfonts}
\usepackage{float}
\usepackage{longtable}
\usepackage{xspace}
\usepackage{stmaryrd}
\usepackage[utf8]{inputenc}
\RequirePackage{hyperref}
\usepackage{graphicx}
\usepackage{color}
\usepackage[linesnumbered,ruled,noline,procnumbered]{algorithm2e}
\usepackage{multirow}

\SetAlgoProcName{Protocol}{List of protocols}

\usepackage{tcolorbox}

\tcbuselibrary{skins,breakable}
\newtcolorbox{functionality}[2][]{%
  enhanced,
  title        = {#2},
  attach boxed title to top left={xshift=+3mm,yshift*=-3mm},
  breakable    = true,
  colback      = blue!5,
  colframe     = blue!35!black,
  fonttitle    = \bfseries,
  colbacktitle = blue!15!white,
  coltitle     = black,
  #1
}

\DeclareMathAlphabet{\mathsl}{OT1}{cmr}{m}{sl}
\DeclareMathAlphabet{\mathsc}{OT1}{cmr}{m}{sc}
\DeclareMathAlphabet{\mathslbf}{OT1}{cmr}{bx}{sl}
\DeclareFontFamily{OT1}{pzc}{}
\DeclareFontShape{OT1}{pzc}{m}{it}%
             {<-> s * [1.150] pzcmi7t}{}
\DeclareMathAlphabet{\mathscript}{OT1}{pzc}{m}{it}

\newtheorem{thm}{Theorem}[section]

\newtheorem{defn}[thm]{Definition}

\newcommand{\figref}[1]{Figure~\ref{#1}}

\newcommand{\Zqm}[2]{\ensuremath{\mathbb{Z}_q^{#1\times #2}}\xspace}

\newcommand{\authnote}[2]{
\ifnum\authnotes=1 
  \begin{center}
    \fbox{\begin{minipage}{5.7in}
      \textbf{#1 says:} #2
    \end{minipage}}
  \end{center} 
\fi
}

\def\getsr{\stackrel{{\scriptscriptstyle\$}}{\leftarrow}}

\newcommand{\plr}{\ensuremath{\pi_{\mathsf{LR-Training}}}\xspace}

\newcommand{\pmul}{\ensuremath{\pi_{\mathsf{DM}}}\xspace}
\newcommand{\pmmul}{\ensuremath{\pi_{\mathsf{DMM}}}\xspace}
\newcommand{\pip}{\ensuremath{\pi_{\mathsf{IP}}}\xspace}

\newcommand{\pdecomp}{\ensuremath{\pi_{\mathsf{decomp}}}\xspace}
\newcommand{\pdecompopt}{\ensuremath{\pi_{\mathsf{decompOPT}}}\xspace}
\newcommand{\ptrunc}{\ensuremath{\pi_{\mathsf{trunc}}}\xspace}

\newcommand{\ptwotoq}{\ensuremath{\pi_{\mathsf{2to2^\lambda}}}\xspace}

\newcommand{\flr}{\ensuremath{\mathcal{F}_{\mathsf{LR-Training}}}\xspace}
\newcommand{\fconv}{\ensuremath{\mathcal{F}_{\mathsf{2to2^\lambda}}}\xspace}

\newcommand{\frho}{\ensuremath{\mathcal{F}_{\rho}}\xspace}
\newcommand{\fmmul}{\ensuremath{\mathcal{F}_{\mathsf{DMM}}}\xspace}

\newcommand{\fdecomp}{\ensuremath{\mathcal{F}_{\mathsf{decomp}}}\xspace}
\newcommand{\fti}[1]{\ensuremath{\mathcal{F}^{\mathcal{D}_{#1}}_{\mathsf{TI}}}\xspace}

\newcommand{\s}{\ensuremath{\mathcal{S}}\xspace}
\newcommand{\env}{\ensuremath{\mathcal{Z}}\xspace}
\newcommand{\adv}{\ensuremath{\mathcal{A}}\xspace}
\newcommand{\F}{\ensuremath{\mathcal{F}}\xspace}

\newcommand{\shareq}[1]{\ensuremath{\llbracket{#1}\rrbracket_{_q}}\xspace}
\newcommand{\sharetwo}[1]{\ensuremath{\llbracket{#1}\rrbracket_{_2}}\xspace}

\begin{document}

\title{High Performance Logistic Regression\\ for Privacy-Preserving Genome Analysis}

\author{Martine De Cock, Rafael Dowsley, Anderson C. A. Nascimento, \\Davis Railsback, Jianwei Shen, Ariel Todoki
\thanks{Martine De Cock, Anderson C. A. Nascimento, Davis Railsback, Jianwei Shen and Ariel Todoki are with the School of Engineering and Technology, University of Washington Tacoma. Emails: \{mdecock, andclay, drail, sjwjames, atodoki\}@uw.edu.}
\thanks{Martine De Cock is a Guest Professor at Ghent University.}
\thanks{Rafael Dowsley is with the Department of Computer Science, Bar-Ilan University, Israel. Email: rafael@dowsley.net. He is supported by the BIU Center for Research in Applied Cryptography and Cyber Security in conjunction with the Israel National Cyber Bureau in the Prime Minister's Office.}
}
\maketitle

\begin{abstract} 
In biomedical applications, valuable data is often split between owners who cannot openly share the data because of privacy regulations and concerns. Training Machine Learning models on the joint data without violating privacy is a major technology challenge that can be addressed by combining techniques from Machine Learning and cryptography.
When collaboratively training Machine Learning models with the cryptographic technique named secure Multi-Party Computation, the price paid for keeping the data of the owners private is an increase in computational cost and runtime. A careful choice of Machine Learning techniques, algorithmic and implementation optimizations are a necessity to enable practical secure Machine Learning over distributed data sets. Such optimizations can be tailored to the kind of data and Machine Learning problem at hand. 

Our setup involves secure Two-Party Computation protocols, along with a trusted initializer that distributes correlated randomness to the two computing parties. We use a gradient descent based algorithm for training a logistic regression model, and we break down the algorithm into corresponding cryptographic protocols. Our main contributions are a new protocol for computing the activation function that requires neither secure comparison protocols nor Yao's garbled circuits, and a series of cryptographic engineering optimizations to improve the performance. To the best of our knowledge, we present the fastest existing secure Multi-Party Computation implementation for training logistic regression models on high dimensional genome data distributed across a local area network.

For our largest gene expression data set, we train a model that requires over 7 billion secure multiplications; the training completes in about 26.90 seconds in a local area network. The implementation in this work is a further optimized version of the implementation with which we won first place in Track 4 of the iDASH 2019 secure genome analysis competition.
 \end{abstract}

\IEEEpeerreviewmaketitle

\section{Background}
\subsection{Introduction}
Machine Learning (ML) has many applications in the biomedical domain, such as medical diagnosis and personalized medicine. Biomedical data sets are typically characterized by high dimensionality, i.e.~a high number of features such as lab test results or gene expression values, and low sample size, i.e.~a small number of training examples corresponding to e.g.~patients or tissue samples. Adding to these challenges, valuable training data is often split between parties (\textit{data owners}) who cannot openly share the data because of privacy regulations and concerns. Due to these concerns, privacy-preserving solutions, using techniques such as secure Multi-Party Computation (MPC), become important so that this data can still be used to train ML models, perform a diagnosis, and in some cases even derive genomic diagnoses \cite{jagadeesh2017deriving}.

We tackle the problem of training a binary classifier on high dimensional gene expression data held by different data owners, while keeping the training data private. This work is directly inspired by Track 4 of the iDASH 2019 secure genome analysis competition\footnote{http://www.humangenomeprivacy.org/2019/\allowbreak competition-tasks.html, accessed on Jan 19, 2020}. The iDASH competition is a yearly international competition for participants to create and implement privacy-preserving protocols for applications with genomic data. The goal is in evaluating the best-known secure methods
and advancing new techniques to solve real-world problems in handling genomic data. In the 2019 edition there were a total of four different tracks, where Track 4 invited participants to design MPC solutions for collaborative training of ML models originating from multiple data owners. One of the Track 4 competition data sets consists of 470 training examples (records) with 17,814 numeric features, while the other consists of 225 training examples with 12,634 numeric features. An initial 5-fold cross-validation analysis in the clear, i.e.~without any encryption, indicated that in both cases logistic regression (LR) models are capable of yielding the level of prediction accuracy expected in the competition, prompting us to investigate MPC-based protocols for secure LR training.

\begin{figure*}[h!]
    \centering
    \includegraphics[width=0.75\textwidth]{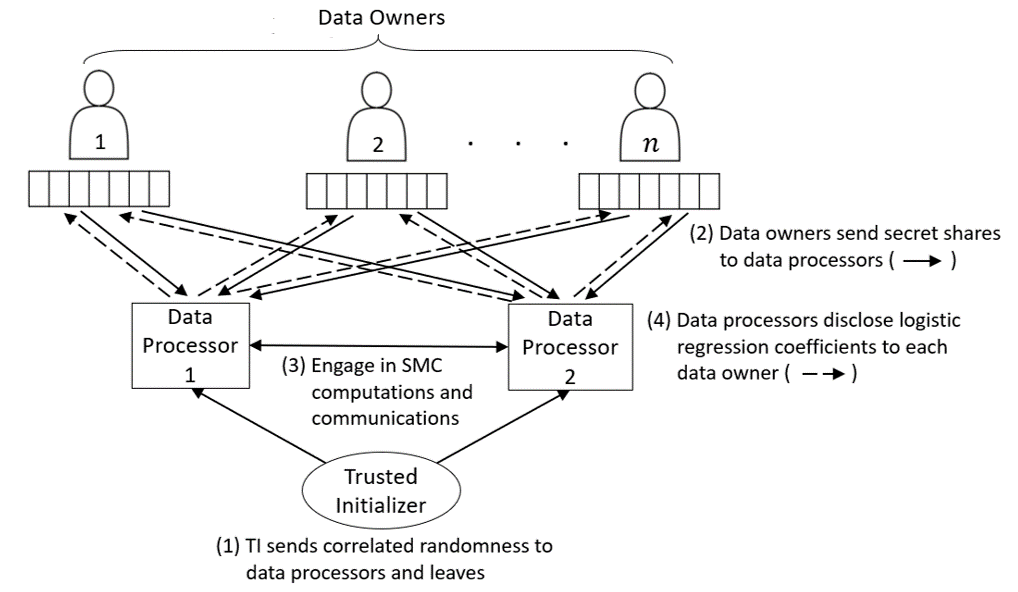}
    \caption{Overview of MPC based secure logistic regression (LR) training.
    Each of $n$ data owners secret shares their own training data between two data processors. The data processors engage in computations and communications to train a ML model, which is at the end revealed to the data owners.}
    \label{fig:diagram}
\end{figure*}

The competition requirements implied the existence of multiple data owners who each send their training example(s) in an encrypted or secret shared form to \textit{data processors} (computing nodes), as illustrated in Figure~\ref{fig:diagram}. The \textit{honest-but-curious}
data processors are not to learn anything about the data as they engage in computations and communications with each other. At the end, they disclose the trained classifier -- in our case, the coefficients of the LR model -- to the data owners. Since the data processors cannot learn anything about the values in the data set, this implies that our protocol is applicable in a wide range of scenarios, independently of how the original data is split by ownership. Our protocol works in scenarios where the data is horizontally partitioned, i.e. when each data owner has different records of the data, such as data belonging to different patients. It also works in scenarios where the data is vertically partitioned, i.e. when each data owner has different features of the data, such as the expression values for different genes.

The main novelty points of our solution for private LR training over a distributed data set are: (i) a new protocol for securely computing the activation function that avoids the use of full-fledged secure comparison protocols; (ii) a novel method for bit decomposing secret shared integers and bundling their instantiations; and (iii) several cryptographic engineering enhancements that together with the novel protocol for the activation function gave us the fastest privacy-preserving LR implementation in the world when run in local area networks (LANs). In summary, we designed a concrete solution for fast secure training of a binary classifier over gene expression data that meets the strict security requirements of the iDASH 2019 competition. For our largest data set, we train a model that requires over 7 billion secure multiplications and the training completes in about 26.9 seconds in a LAN. 

This paper significantly expands over a preliminary version of this result \cite{PRIML2019}, presented at a workshop without formal proceedings. In this version we have a formal description of all protocols, security proofs and improved running times.

We first discuss below our work as compared to others. In the Section \textbf{Methods}, we present preliminary information on MPC, describe the secure subprotocols that are building blocks for our secure LR training protocol, and finally describe the protocol itself. In the Section \textbf{Results} we describe details of our implementation and runtime results for the overall protocol and microbenchmarks for our secure activation function protocol. We experimentally compare our solution with the state-of-the-art SecureML approach \cite{mohassel2017secureml}, demonstrating substantial runtime improvements.
In the Section \textbf{Discussion}, we note possible future work to improve and extend our results, and finally in the Section \textbf{Conclusions} we present our summary remarks.

\subsection{Related Work}
A variety of efforts have previously been made to train LR classifiers in a privacy-preserving way.

One scenario that was considered in previous works \cite{bonte2018privacy,chen2018logistic, kim2018logistic} is the setting in which a data owner holds the data while another party (the data processor), such as a cloud service, is responsible for the model training. These solutions usually rely on homomorphic encryption, with the data owner encrypting and sending their data to the data processor who performs computations on the encrypted data without having to decrypt it. 

When the data is held by multiple data owners, they can either execute an MPC protocol among themselves to train the model, or delegate the computation to a set of data processors that run a MPC protocol. It is the latter setting that we follow in this paper.

\begin{figure*}[h!]
    \centering
    \includegraphics[width=0.75\textwidth]{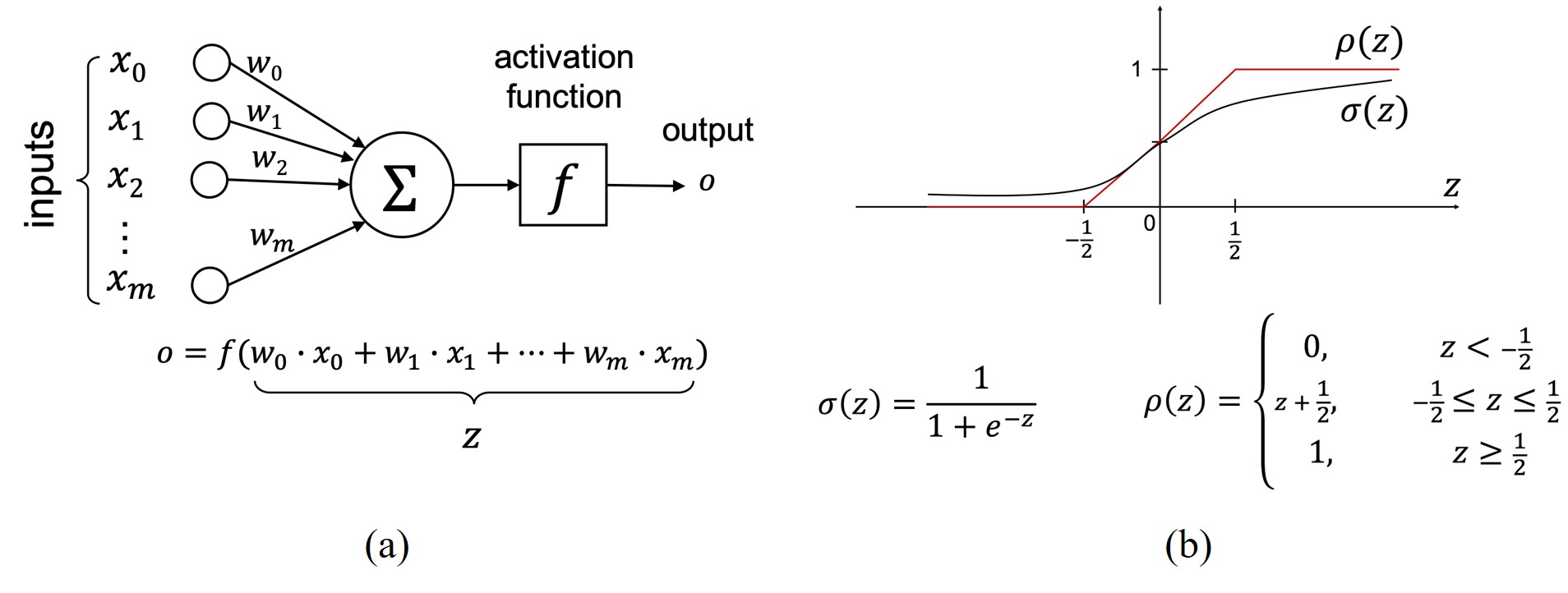}
    \caption{Architecture: (a) Neuron; (b) Approximation of sigmoid activation function $\sigma$ by clipped ReLu $\rho$}
    \label{fig:architecture}
\end{figure*}

Existing MPC approaches to secure LR differ in the numerical optimization algorithms used for LR training and in the cryptographic primitives leveraged \cite{el2012secure, mohassel2017secureml, nardi2012achieving,xie2016privlogit}. The SPARK protocol \cite{el2012secure} uses additive homomorphic encryption (Paillier cryptosystem) and uses Newton-Raphson as the numerical optimization algorithm to find the values of the weights that maximize the log-likelihood. The SPARK protocol can use the actual logistic function without approximating it at the cost of the plaintext data being horizontally partitioned and seen by the data processors. The two protocols from \cite{nardi2012achieving} rely on the Newton-Raphson method, both approximate the logistic function, and both use additive secret sharing. The first protocol includes the use of Yao's garbled circuits to compute the approximation of the logistic function, while the second protocol uses a Taylor approximation and Euler's method. The PrivLogit method \cite{xie2016privlogit} uses Yao's garbled circuits and Paillier encryption; their protocol uses the Newton-Raphson method and a constant Hessian approximation to speed up computation. However, this protocol relies on the plaintext data being horizontally partitioned and seen by the data processors, which, like the work in \cite{el2012secure}, would not align with the iDASH 2019 competition requirements. We also point out a protocol secure against active adversaries from SecureNN \cite{wagh2019securenn} for computing a ReLu. While we compute a different function (clipped ReLu), we share a similar idea that using the most significant bit of an input can tell us the output of the function.

The work closest to ours is SecureML  \cite{mohassel2017secureml}, which was the fastest protocol for privately training LR models based on secure MPC prior to our work. SecureML separates the data owners from the data processors, and uses mini-batch gradient descent. The main novelty points of SecureML are a clipped ReLu activation function, a novel truncation protocol, and a combination of garbled circuits and secret sharing based MPC in order to obtain a good trade-off between communication, computation and round complexities. The SecureML protocol is evaluated on a data set with up to 5,000 features, while -- to the best of our knowledge -- the existing runtime evaluation of all other approaches for MPC based LR training is limited to 400 features or less  \cite{el2012secure,nardi2012achieving,xie2016privlogit}. Like our solution, the SecureML protocol is split into an offline and online phase (the offline phase can be executed before the inputs are known and is responsible for generating multiplication triples). The SecureML solution is based on two servers, while our solution is based on three servers, namely a party who pre-computes so-called multiplication triples in the off-line stage, and two parties who actively compute the final result. If we exclude the preprocessing/off-line stage from SecureML and exclude the pre-distribution of triples in our solution, we are left with protocols that work in exactly the same setting. We compare the runtime of both solutions in the Section \textbf{Results}, showing that our implementation is substantially faster.

A preliminary version of this work appeared in a workshop without formal proceedings \cite{PRIML2019}. This paper is a substantially longer and detailed description that includes security proofs, detailed comparison with the state-of-the-art, and improved running times.

%Despite our solution being based on three servers and the SecureML solution being based on two servers, a comparison between these protocols is meaningful. One of the parties in our solution is used solely to pre-compute so-called multiplication triples. The other two parties, in our solution, are the ones actively computing the final result.   

\section{Methods}

\subsection{Logistic Regression}
Logistic regression is a common Machine Learning algorithm for binary classification. The training data $D$ consists of training examples $d = (\boldsymbol{x}_d,t_d)$ in which $\boldsymbol{x}_d=\langle x_{d,1},x_{d,2},...,x_{d,m}\rangle$ is an $m$-dimensional numerical vector, containing the values of $m$ input attributes for example $d$, and $t_d\in \{0,1\}$ is the ground truth class label. Each $x_{d,i}$ for $i \in \{ 1,2,...,m\}$ is a real number value.

As illustrated in Figure \ref{fig:architecture}(a), we train a neuron to map the $\boldsymbol{x}_d$'s to the corresponding $t_d$'s, correctly classifying the examples. The neuron computes a weighted sum of the inputs (the values of the weights are learned during training) and subsequently applies an activation function to it, to arrive at the output $o_d = f(w_0\cdot x_{d,0}+w_1\cdot x_{d,1}+\cdots+w_n\cdot x_{d,n})$, which is interpreted as the probability that the class label is 1. Note that, as is common in neural network training, we extend the input attribute vector with a dummy feature $x_{d,0}$ which has value 1 for all $\boldsymbol{x}_d$'s. The traditionally used activation function for LR is the sigmoid function $\sigma(z)=\frac{1}{1+e^{-z}}$. Since the sigmoid function $\sigma$ requires division and evaluation of an exponential function, which are expensive operations to perform in MPC, we approximate it with the activation function $\rho$ from \cite{mohassel2017secureml}, which is shown in Figure \ref{fig:architecture}(b). 

\SetAlFnt{\normalsize\normalfont}
\begin{algorithm}[h]
%\SetAlgoLined
\SetKwInOut{Input}{Input}
\SetKwInOut{Output}{Output}
\SetKwFor{For}{for}{do}{}
\SetKwFor{Until}{until}{do}{}
\SetKwFor{ForE}{for each}{do}{}

\Input{A set $D$ with training examples $(\boldsymbol{x}_d,t_d)$; a learning rate $\eta$}
\Output{Weights $w_i$ that minimize the sum of squared errors over the training data}
\For{ $i \leftarrow 0$ \KwTo $m$}{
    $w_i \leftarrow 0$
}
\Until{termination condition is met}{
    \For{$i \leftarrow 0$ \KwTo $m$}{
        $\Delta w_i \leftarrow 0$
    }
    \ForE{$(\boldsymbol{x}_d,t_d)$ in $D$}{
        $o_d \leftarrow \rho(w_0 \cdot x_{d,0} + w_1 \cdot x_{d,1} + \ldots + w_m \cdot x_{d,m})$\\
        \For{$i \leftarrow 0$ \KwTo $m$}{
            $\Delta w_i \leftarrow \Delta w_i + \eta (t_d-o_d) x_{d,i}$
        }
    }
    \For{$i \leftarrow 0$ \KwTo $m$}{
        $w_i \leftarrow w_i + \Delta w_i$
    }
}
\KwRet{$w_0, \ldots, w_m$}
 \caption{Full Gradient Descent\label{alg:gradientdescent}}
\end{algorithm}

For training, we use the full gradient descent based algorithm shown in Algorithm \ref{alg:gradientdescent} to learn the weights for the LR model. On line 3, we choose not to use \textit{early stopping}\footnote{This is a technique that uses a metric, such as the accuracy on a held-out validation data set, to check when a model starts to overfit and will then stop training at that point. } because in that case the number of iterations would depend on the values in the training data, hence leaking information \cite{nardi2012achieving}. Instead, we use a fixed number of iterations during training.

\subsection{Our scenario}

In the scenario considered in this work the data is not held by a single party that performs all the computation, but distributed by the data owners to the data processors in such way that each data processor does not have any information about the data in the clear. Nevertheless, the data processors would still like to compute a LR model without leaking any other information about the data used for the training. To achieve this goal, we will use techniques from MPC. 

Our setup is illustrated in Figure \ref{fig:diagram}. We have multiple data owners 
who each hold disjoint parts of the data that is going to be used for the training. This is the most general approach and covers the cases in which the data is horizontally partitioned (i.e. for each training sample $d=(\boldsymbol{x}_d,t_d)$, all the data for $d$ is held by one of the data owners), vertically partitioned (for each feature, the values of that feature for all training samples are held by one of the data owners), and even arbitrary partitions. There are
two data processors who collaborate to train a LR model using secure MPC protocols, and a trusted initializer (TI) that predistributes correlated randomness to the data processors in order to make the MPC computation more efficient. The TI is not involved in any other part of the execution, and does not learn any data from the data owners or data processors.

We next present the security model that is used and several secure building blocks, so that afterwards we can combine them in order to obtain a secure LR training protocol.

\subsection{Security Model}\label{secmodel}

The security model in which we analyze our protocol is the Universal Composability (UC) framework \cite{FOCS:Canetti01} as it provides the strongest security and composability guarantees and is the gold standard for analyzing cryptographic protocols nowadays. Here we will only give a short overview of the UC framework (for the specific case of two-party protocols), and refer interested readers to the book of Cramer et al. \cite{CDN2015} for a detailed explanation. 

The main advantage of the UC framework is that the UC composition theorem guarantees that any protocol proven UC-secure can also be securely composed with other copies of itself and of other protocols (even with arbitrarily concurrent executions) while preserving its security. Such guarantee is very useful since it allows the modular design of complex protocols, and is a necessity for protocols executing in complex environments such as the Internet.

The UC framework first considers a real world scenario in which the two protocol participants (the data processors from Figure \ref{fig:diagram}, henceforth denoted Alice and Bob) interact between themselves and with an adversary $\adv$ and an environment $\env$ (that captures all activity external to the single execution of the protocol that is under consideration). The environment $\env$ gives the inputs and gets the outputs from Alice and Bob. The adversary $\adv$ delivers the messages exchanged between Alice and Bob (thus modeling an adversarial network scheduling) and can corrupt one of the participants, in which case he gains the control over it. In order to define security, an ideal world is also considered. In this ideal world, an idealized version of the functionality that the protocol is supposed to perform is defined. The ideal functionality $\F$ receives the inputs directly from Alice and Bob, performs the computations locally following the primitive specification and delivers the outputs directly to Alice and Bob. A protocol $\pi$ executing in the real world is said to UC-realize functionality $\F$ if for every adversary $\adv$ there exists a simulator $\s$ such that no environment $\env$ can distinguish between: (1) an execution of the protocol $\pi$ in the real world with participants Alice and Bob, and adversary $\adv$; (2) and an ideal execution with dummy parties (that only forward inputs/outputs), $\F$ and $\s$.

This work like the vast majority of the privacy-preserving machine learning protocols in the literature considers honest-but-curious, static adversaries. In more detail, the adversary chooses the party that he wants to corrupt before the protocol execution and he also follows the protocol instructions (but tries to learn additional information). We consider the trusted initializer model, in which a trusted initializer functionality $\fti{}$ pre-distributes correlated randomness to Alice and Bob.\footnote{Using a setup assumption, like the trusted initializer, in the MPC protocol is a necessity in order to get UC-security \cite{C:CanFis01,STOC:CLOS02}. Other possible setup assumption to achieve UC-security include: a common reference string \cite{C:CanFis01,STOC:CLOS02,C:PeiVaiWat08}, the availability of a public-key infrastructure \cite{FOCS:BCNP04}, the random oracle model \cite{TCC:HofMul04,EPRINT:BDDMN17b}, the existence of noisy channels between the parties \cite{SBSEG:DMN08,JIT:DGMN13}, and the availability of tamper-proof hardware \cite{EC:Katz07,ICITS:DowMulNil15}.} A trusted initializer has been often used to enable highly efficient solutions both in the context of privacy-preserving machine learning \cite{AISec:CDNN15,david2015efficient,fritchman2018,IEEETDSC:CDHK+17,NeurIPS2019} as well as in other applications, e.g., \cite{r99,dowsley2010two,IEICE:DMOHIN11,ishai2013power,IJIS:TNDMIHO15,IEEEIFS:DDGM+16}. 

\begin{functionality}{Functionality $\fti{}$}
$\fti{}$ is parametrized by an algorithm $\mathcal{D}$ for sampling the correlated randomness. Upon initialization, run $(D_A, D_B) \getsr \mathcal{D}$, and deliver $D_A$ to Alice and $D_B$ to Bob.
\end{functionality}

\textbf{Simplifications:} In our proofs the simulation strategy is simple and will be described briefly: all the messages look uniformly random from the recipient's point of view, except for the messages that open a secret shared value to a party, but these ones can be easily simulated using the output of the respective functionalities. Therefore a simulator $\s$, having the leverage of being able to simulate the trusted initializer functionality $\fti{}$ in the ideal world, can easily perform a perfect simulation of a real protocol execution; therefore making the real and ideal worlds indistinguishable for any environment $\env$. In the ideal functionalities the messages are public delayed outputs, meaning that the simulator is first asked whether they should be delivered or not (this is due to the modeling that the adversary controls the network scheduling). This fact as well as the session identifications are omitted from our functionalities' descriptions for the sake of readability.   

\subsection{Secret Sharing Based Secure Multi-Party Computation}

Our MPC solution is based on additive secret sharing over a ring $\mathbb{Z}_{q}$ $=$ $\{0,1,\ldots,q-1\}$. When secret sharing a value $x \in \mathbb{Z}_{q}$, Alice and Bob receive shares $x_A$ and $x_B$, respectively, that are chosen uniformly at random in $\mathbb{Z}_{q}$ with the constraint that $x_A + x_B = x \mod q$. We denote the pair of shares by $[\![x]\!]_q$. All computations are modulo $q$ and the modular notation is henceforth omitted for conciseness. Note that no information of the secret value $x$ is revealed to either party holding only one share. The secret shared value can be revealed/opened to each party by combining both shares. Some operations on secret shared values can be computed locally with no communication. Let $[\![x]\!]_q$, $[\![y]\!]_q$ be secret shared values and $c$ be a constant. Alice and Bob can perform the following operations locally:
\begin{itemize}
  \item Addition ($z=x+y$): Each party locally adds its local shares of $x$ and $y$ in order to obtain a share of $z$. This will be denoted by 
  $[\![z]\!]_q \leftarrow[\![x]\!]_q+[\![y]\!]_q$.
  \item Subtraction ($z=x-y$): Each party locally subtracts its local share of $y$ from that of $x$ in order to obtain a share of $z$. This will be denoted by $[\![z]\!]_q\leftarrow[\![x]\!]_q-[\![y]\!]_q$.
  \item Multiplication by a constant ($z=cx$): Each party multiplies its local share of $x$ by $c$ to obtain a share of $z$. This will be denoted by 
  $[\![z]\!]_q\leftarrow c[\![x]\!]_q$
  \item Addition of a constant ($z=x+c$): Alice adds $c$ to her share $x_A$ of $x$ to obtain $z_A$, while Bob sets $z_B=x_B$. This will be denoted by $[\![z]\!]_q\leftarrow[\![x]\!]_q + c$.
\end{itemize}

The secure multiplication of secret shared values (i.e., $z=xy$) cannot be done locally and  involves communication between Alice and Bob. To obtain an efficient secure multiplication solution, we use the multiplication triples technique that was originally proposed by Beaver \cite{beaver1997commodity}. We use a trusted initializer to pre-distribute the multiplication triples (which are a form of correlated randomness) to Alice and Bob.  We use the same protocol $\pmmul$ for secure (matrix) multiplication of secret shared values as in \cite{IEEETDSC:CDHK+17,Dowsley16} and denote by $\pmul$ the protocol for the special case of multiplication of scalars and $\pip$ for the inner product. As shown in \cite{IEEETDSC:CDHK+17} the protocol $\pmmul$ (described in Protocol \ref{prot:distmult}) UC-realizes the distributed matrix multiplication functionality $\fmmul$ in the trusted initializer model.

\begin{functionality}{Functionality $\fmmul{}$}
$\fmmul$ runs with Alice and Bob and is parametrized by the size $q$ of the ring $\mathbb{Z}_q$ and the dimensions $(i, j)$ and $(j, k)$ of the matrices.\\
\\
\textbf{Input:} Upon receiving a message from Alice/Bob with its shares of $\shareq{X}$ and $\shareq{Y}$, verify if the share of $X$ is in $\Zqm{i}{j}$ and the share of $Y$ is in $\Zqm{j}{k}$.
If it is not, abort. Otherwise, record the shares, ignore any subsequent message from that party and
inform the other party about the receipt.\\

\textbf{Output:} Upon receipt of the shares from both parties, reconstruct $X$ and $Y$ from 
the shares, compute $Z=X Y$ and create a secret sharing $\shareq{Z}$ to distribute to Alice and Bob: a corrupt party fixes its share of the output to any chosen matrix and the shares of the uncorrupted parties are then created by picking uniformly random values subject to the correctness constraint.
\end{functionality}

\begin{procedure}[hbt!]
     \SetKwInOut{Input}{Input}
     \SetKwInOut{Output}{Output}

     \Input{$[\![X]\!]_q$, $[\![Y]\!]_q$}
     \Output{$[\![Z]\!]_q$ such that $Z=XY$}

    The protocol is parametrized by the size $q$ of the ring $\mathbb{Z}_q$ and the dimensions $(i, j)$ and $(j, k)$ of the matrices. The trusted initializer chooses uniformly random $U$ and $V$ in $\mathbb{Z}_q^{i \times j}$ and $\mathbb{Z}_q^{j \times k}$, respectively, computes $W=UV$ and pre-distributes secret sharings $[\![U]\!]_q, [\![V]\!]_q, [\![W]\!]_q$ to Alice and Bob. 

    Alice and Bob locally compute $[\![D]\!]_q \gets [\![X]\!]_q-[\![U]\!]_q$ and $[\![E]\!]_q \gets [\![Y]\!]_q-[\![V]\!]_q$, and then open $D$ and $E$.\\
    
    Alice and Bob locally compute $[\![Z]\!]_q \gets [\![W]\!]_q + E [\![U]\!]_q + D [\![V]\!]_q +DE$.\\
    \KwRet{$[\![Z]\!]_q$}

     \caption{Secure Distributed Matrix Multiplication Protocol() $\pmmul$\label{prot:distmult}}

\end{procedure}

\begin{figure*}[h!]
    \centering
    \includegraphics[width=0.75\textwidth]{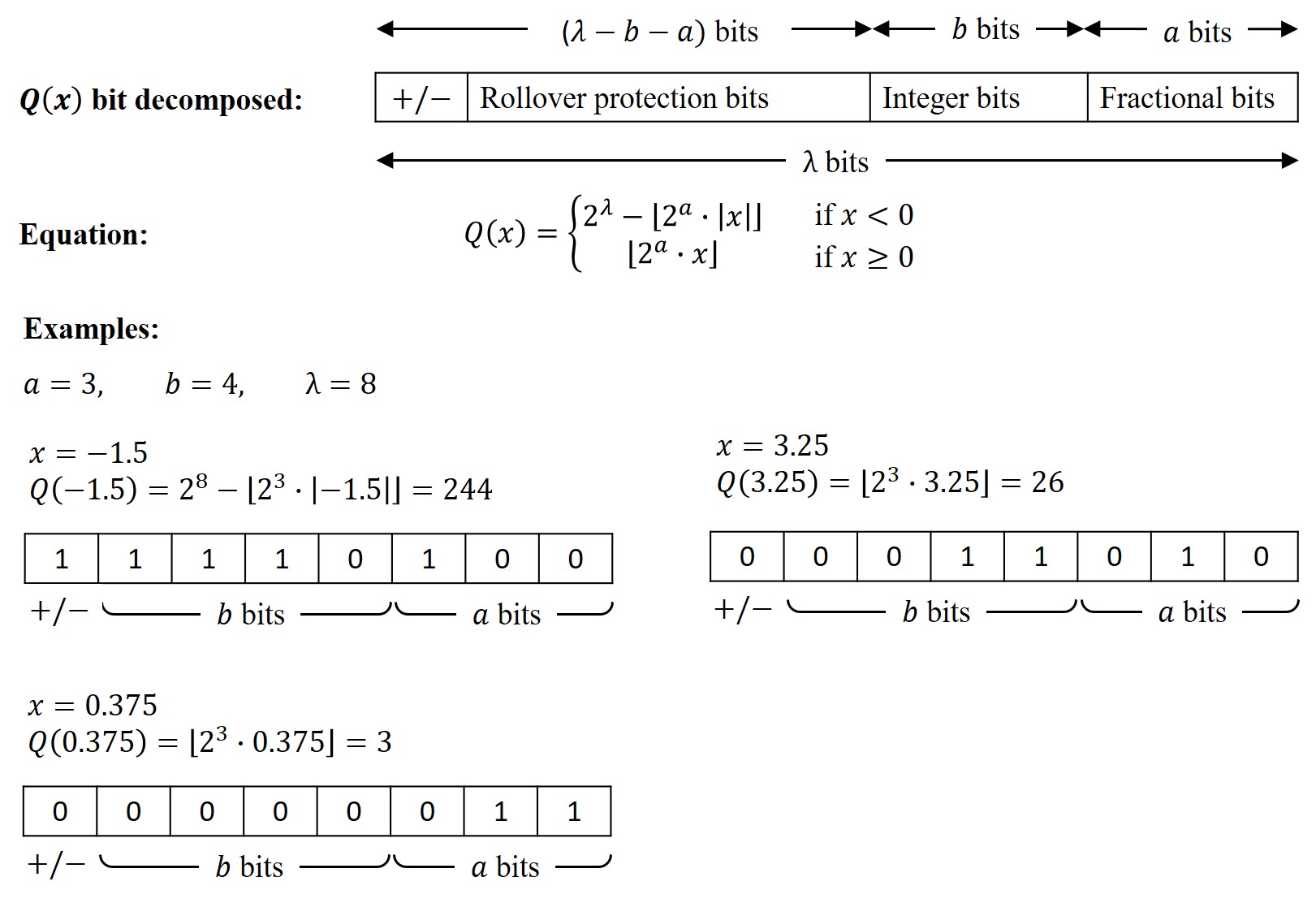}
    \caption{Fixed-Point Representation: Register map of fixed-point representation of numbers shared over $\mathbb{Z}_{2^\lambda}$ with examples.}
    \label{fig:registermap}
\end{figure*}

\subsection{Converting to Fixed-Point Representation}

Each data owner initially needs to convert their training data to integers modulo $q$ so that they can be secret shared. As illustrated in Figure  \ref{fig:registermap}, each feature value $x \in \mathbb{R}$ is converted into a fixed point approximation of $x$ using a two's complement representation for negative numbers. We define this new value as $Q(x) \in \mathbb{Z}_q$. This conversion is shown in Equation (\ref{eq:fixedpoint}): 

\begin{equation}
\label{eq:fixedpoint}
Q(x) = 
\begin{cases} 
      2^\lambda - \left \lfloor{ 2^a \cdot |x|  }\right \rfloor & \mbox{if\ } x < 0 \\
      \left \lfloor{ 2^a \cdot x  }\right \rfloor & \mbox{if\ } x \geq 0
\end{cases}
\end{equation}

Specifically, when we convert $Q(x)$ into its bit representation, we define the first $a$ bits from the right to hold the fractional part of $x$, and the next $b$ bits to represent the non-negative integer part of $x$, and the most significant bit (MSB) to represent the sign (positive or negative). We define $\lambda$ to represent the total number of bits such that the ring size $q$ is defined as $q=2^\lambda$. It is important to choose a $\lambda$ that is large enough to represent the largest number $x$ that can be produced during the LR protocol, and therefore $\lambda$ should be chosen to be at least $2(a+b)$ (see Truncation). 
It is also important to choose a $b$ that is large enough to represent the maximum possible value of the integer part of all $x$'s (this is dependent on the data). This conversion and bit representation is shown in Figure \ref{fig:registermap}.

\subsection{Truncation}
When multiplying numbers that were converted into a fixed point representation with $a$ fractional bits, the resulting product will end up with $a$ more bits representing the fractional part. For example, a fixed point representation of $x$ and $y$, for $x, y > 0$, is $x\cdot2^a$ and $y\cdot2^a$, respectively. The multiplication of both these terms results in $xy\cdot2^{2a}$, showing that now $2a$ bits are representing the fractional part, which we must scale back down to $xy\cdot2^a$ to do any further computations. In our solution, we use the two-party local truncation protocol for fixed point representations of real numbers proposed in \cite{mohassel2017secureml} that we will refer to as $\ptrunc$. It does not involve any messages between the two parties, each party simply performs an operation on its own local share. This protocol almost always incurs an error of at most a bit flip in the least-significant bit. However, with probability $2^{a  +1-\lambda}$, where $a$ is the number of fractional bits, the resulting value is completely random. 

When this truncation protocol is performed on increasingly large data sets (in our case we run over 7 billion secure multiplications), the probability of an erroneous truncation becomes a real issue -- an issue not significant in previous implementations. There are two phases in which truncation is performed: (1) when computing the dot product (inner product) of the current weights vector with a training example in line 7 of Algorithm \ref{alg:gradientdescent}, and (2) when the weight differentials ($\Delta w_i$) are adjusted in line 9 of Algorithm \ref{alg:gradientdescent}. If a truncation error occurs during (1), the resulting erroneous value will be pushed into a reasonable range by the activation function and incur only a minor error for that round. If the error occurs during (2), an element of the weights vector will be updated to a completely random ring element and recovery from this error will be impossible. To mitigate this in experiments, we make use of 10-12 bits of fractional precision with a ring size of 64 bits, making the probability of failure $\frac{1}{2^{53}} < p < \frac{1}{2^{51}}$. The number of truncations that need to be performed is also reduced in our implementation by waiting to perform truncation until it is absolutely required. For instance, instead of truncating each result of multiplication between an attribute and its corresponding weight, a single truncation can be performed at the end of the entire dot product.  

Additional error is incurred on the accuracy by the fixed point representation itself. Through cross-validation with an in-the-clear implementation, we determined that 12 bits of fractional precision provide enough accuracy to make the output accuracy indistinguishable between the secure version and the plaintext version.

\subsection{Conversion of Sharings}

For efficiency reasons, in some of the steps for securely computing the activation function we use secret sharings over $\mathbb{Z}_2$, while in others we 
use secret sharings over $\mathbb{Z}_{2^{\lambda}}$. Therefore we need to be able to convert between the two types of secret sharings.

We use the two-party protocol from \cite{IEEETDSC:CDHK+17}
for performing the bit-decomposition of a secret-shared value $[\![x]\!]_{2^{\lambda}}$ to shares $\sharetwo{x_i}$, where $x_\lambda \cdots x_1$ is the binary representation of $x$.
It works like the ripple carry adder arithmetic circuit based on the insight that the difference between the sum of the two additive shares held by the parties and an ``XOR-sharing'' of that sum is the carry vector. As proven in \cite{IEEETDSC:CDHK+17}, the bit-decomposition protocol $\pdecomp$ (described in Protocol \ref{prot:decomp}) UC-realizes the bit-decomposition functionality $\fdecomp$.

\begin{functionality}{Functionality $\fdecomp$}
$\fdecomp$ runs with Alice and Bob and is parametrized by the bit-length $\lambda$ of the value $x$ being converted from additive sharings $[\![x]\!]_{2^{\lambda}}$ in $\mathbb{Z}_{2^{\lambda}}$ to additive bitwise sharings $\sharetwo{x_i}$ in $\mathbb{Z}_2$ such that $x=x_\lambda \cdots x_1$.\\
\\
\textbf{Input:} Upon receiving a message from Alice or Bob with its share of $[\![x]\!]_{2^{\lambda}}$, record the share, ignore any subsequent messages from that party and inform the other party about the receipt.\\
\\
\textbf{Output:} Upon receipt of the inputs from both parties, reconstruct the value $x=x_\lambda \cdots x_1$ from the shares, and for $i\in \{1,\ldots,\lambda\}$ distribute new sharings $\sharetwo{x_i}$ of the bit $x_i$. Before the output deliver, the corrupt party fix its shares of the output to any desired value. The shares of the uncorrupted parties are then created by picking uniformly random values subject to the correctness constraints.
\end{functionality}

\begin{procedure}[hbt!]
     \SetKwInOut{Input}{Input}
     \SetKwInOut{Output}{Output}

     \Input{$[\![x]\!]_{2^{\lambda}}$}
     \Output{$\sharetwo{x_i}$, where $x_\lambda \cdots x_1$ is the binary representation of $x$.}

    All distributed multiplications are over $\mathbb{Z}_2$ and the required correlated randomness is pre-distributed by the trusted initializer.\\
    
    Let $a$ denote Alice's share of $x$, which corresponds to a bit string $a_\lambda \ldots a_1$. Similarly, let $b$ denote Bob's share of $x$, which corresponds to a bit string $b_\lambda \ldots b_1$.  Define the secret sharings $\sharetwo{y_i}$ as the pair of shares $(a_i,b_i)$ for $y_i=a_i+b_i \mod{2}$, $\sharetwo{a_i}$ as $(a_i,0)$ and $\sharetwo{b_i}$ as $(0,b_i)$.\\
    
    Compute $\sharetwo{c_1} \gets \sharetwo{a_1} \sharetwo{b_1}$ and set $\sharetwo{x_1} \gets \sharetwo{y_1}$.\\
    
    \For{$i\gets2$ \KwTo $\lambda$}{
        Compute $\sharetwo{d_i} \gets \sharetwo{a_i} \sharetwo{b_i} + 1$.\\
        
        $\sharetwo{e_i} \gets \sharetwo{y_i} \sharetwo{c_{i-1}} + 1$\\
        
        $\sharetwo{c_i} \gets \sharetwo{e_i} \sharetwo{d_i} + 1$\\
        
        $\sharetwo{x_i} \gets \sharetwo{y_i} + \sharetwo{c_{i-1}}$\\
        
    }
    
    \KwRet{$\sharetwo{x_i}$ for $i\in \{1,\ldots,\lambda\}$}
    
     \caption{Secure Two-Party Bit-Decomposition Protocol() $\pdecomp$\label{prot:decomp}}

\end{procedure}

In our implementation we use a highly parallelized and optimized version of the bit-decomposition protocol $\pdecomp$ in order to improve the communication efficiency of the overall solution. The optimizations are described in the Appendix.

The opposite of a secure bit-decomposition is converting from bit sharing to an additive sharing over a larger ring. In our secure activation function protocol, we require securely converting a bit sharing to an additive sharing in $2^\lambda$. This is done using the protocol $\ptwotoq$ from \cite{NeurIPS2019} (described in Protocol \ref{prot:conv}) that UC-realizes the secret sharing conversion functionality $\fconv$.

\begin{functionality}{Functionality $\fconv$}
$\fconv$ is parametrized by the bit-length $\lambda$ of the ring in which the output is shared. \\
\\
\textbf{Input:} Upon receiving a message from Alice/Bob with her/his share of $\sharetwo{x}$, record the share, ignore any subsequent messages from that party and inform the other party about the receipt.\\
 \\
\textbf{Output:} Upon receipt of the inputs from both parties, reconstruct $x$, then create and distribute to Alice and Bob the secret sharing $[\![x]\!]_{2^{\lambda}}$. Before the deliver of the output shares, a corrupt party fix its share of the output to any constant value. In both cases the shares of the uncorrupted parties are then created by picking uniformly random values subject to the correctness constraint.
\end{functionality}

\begin{procedure}[hbt!]
     \SetKwInOut{Input}{Input}
     \SetKwInOut{Output}{Output}

     \Input{$[\![x]\!]_{2}$}
     \Output{$[\![x]\!]_{2^{\lambda}}$}

    All distributed multiplications are over $\mathbb{Z}_{2^{\lambda}}$ and the required correlated randomness is pre-distributed by the trusted initializer.\\
    
    For the input $[\![x]\!]_2$, let $x_A \in \{0,1\}$ denote Alice's share and $x_B \in \{0,1\}$ denote Bob's share.\\
    
    Alice creates a secret sharing $[\![x_A]\!]_{2^{\lambda}}$ by picking uniformly random shares that sum to $x_A$ and delivers Bob's share to him, and Bob proceeds similarly to create $[\![x_B]\!]_{2^{\lambda}}$. \\
    
    $[\![y]\!]_{2^{\lambda}} \gets [\![x_A]\!]_{2^{\lambda}}[\![x_B]\!]_{2^{\lambda}}$\\
    
    $[\![z]\!]_{2^{\lambda}} \gets [\![x_A]\!]_{2^{\lambda}}+[\![x_B]\!]_{2^{\lambda}}-2[\![y]\!]_{2^{\lambda}}$\\
    \KwRet{$[\![z]\!]_{2^{\lambda}}$}
    
     \caption{Secure Secret Sharing Conversion Protocol() $\ptwotoq$\label{prot:conv}}

\end{procedure}

\subsection{Secure Activation Function}

We propose a new protocol that evaluates $\rho$ from Figure \ref{fig:architecture}(b) directly over additive shares and does not require full secure comparisons, which would have been more expensive. Instead of doing straightforward comparisons between $z$, $0.5$ and $-0.5$, we derive the result through checking two things: (i) whether $z'=z+ 1/2$ is positive or negative; (ii) whether $z' \geq 1$. Both checks can be performed without using a full comparison protocol. 

When $z'$ is bit decomposed, the most significant bit is 0 if $z'$ is non-negative and 1 if $z'$ is negative. In fact, if out of the $\lambda$ bits, the $a$ lowest bits are used to represent the fractional component and the $b$ next bits are used to represent the integer component, then the remaining $\lambda-a-b$ bits all have the same value as the most significant bit. We will use this fact in order to optimize the protocol by only performing a partial bit-decomposition and deducting whether $z'$ is positive or negative from the $(a+b+1)$-th bit. 

In the case that $z'$ is negative, the output of $\rho$ is $0$. But, if $z'$ is positive, we need to determine whether $z' \geq 1$ in order to know if the output of $\rho$ should be fixed to $1$ or to $z'$. A positive $z'$ is such that $z' \geq 1$ if and only if at least one of the $b$ bits corresponding to the integer component of $z'$ representation is equal to 1, therefore we only need to analyze those $b$ bits to determine if $z' \geq 1$.

Our secure protocol $\pi_\rho$ is described in Protocol \ref{alg:secureactivation}. The AND operation corresponds to multiplications in $\mathbb{Z}_{2}$. 
By the application of De Morgan's law, the OR operation is performed using the AND and negation operations. The successive multiplications can be optimized to only take a logarithmic number of rounds by using well-known techniques.

The activation function protocol $\pi_\rho$ UC-realizes the activation function functionality $\frho$. The correctness can be checked by inspecting the three possible cases: (i) if $z > 1/2$, then $\mathsf{pos}=1$ and $\mathsf{geq1}=1$ (since at least one of the bits representing the integer component of $z+1/2$ will have a value 1). The output is thus $[\![2^a]\!]_{2^{\lambda}}$ (the fixed-point representation of 1); if $-1/2 \leq z < 1/2$, then $\mathsf{pos}=1$ and $\mathsf{geq1}=0$, and therefore the output will be $[\![z']\!]_{2^{\lambda}}$, which is the fixed-point representation of $z+1/2$; if $z<-1/2$, then $\mathsf{pos}=0$ and the output will be a secret sharing representing zero as expected. The security follows trivially from the UC-security of the building blocks used and the fact that no secret sharing is opened. 

\begin{functionality}{Functionality $\frho$}
\textbf{Input:} Upon receiving a message from Alice/Bob with her/his share of $[\![z]\!]_{2^{\lambda}}$, record the share, ignore any subsequent messages from that party and inform the other party about the receipt.\\
 \\
\textbf{Output:} Upon receipt of the inputs from both parties, reconstruct $z$, compute the result of the activation function $\rho(z)$, and then create and distribute to Alice and Bob the secret sharing $[\![\rho(z)]\!]_{2^{\lambda}}$ (using the fixed-point representation). Before the deliver of the output shares, a corrupt party fix its share of the output to any constant value. In both cases the shares of the uncorrupted parties are then created by picking uniformly random values subject to the correctness constraint.
\end{functionality}

\begin{procedure}[h]
    \SetKwInOut{Input}{Input}
    \SetKwInOut{Output}{Output}

    \Input{ $[\![z]\!]_{2^{\lambda}}  $}
    \Output{$[\![\rho(z)]\!]_{2^{\lambda}}$}
    
    $[\![z']\!]_{2^{\lambda}} \leftarrow [\![z]\!]_{2^{\lambda}}+ 2^{a-1}$ \\
    
    $[\![z'_1]\!]_2, \ldots [\![z'_{a+b+1}]\!]_2 \leftarrow \pdecomp([\![z']\!]_{2^\lambda}\land(2^{a+b+2} - 1))$ \\

     $[\![\mathsf{pos}]\!]_{2^{\lambda}} \leftarrow \ptwotoq( 1 - [\![z'_{a+b+1}]\!]_2 )$ \\
    
    $[\![\mathsf{geq1}]\!]_{2} \leftarrow \bigvee_{i \in \{a+1,...,a+b\}} [\![z'_{i}]\!]_2$ \\
    
    $[\![\mathsf{geq1}]\!]_{2^\lambda} \leftarrow \ptwotoq( [\![\mathsf{geq1}]\!]_{2} ) $ \\
    
    $[\![r]\!]_{2^{\lambda}} \leftarrow 2^a[\![\mathsf{geq1}]\!]_{2^\lambda} + \left(1-[\![\mathsf{geq1}]\!]_{2^\lambda}\right)[\![z']\!]_{2^{\lambda}}$ \\

    $[\![\rho(z)]\!]_{2^{\lambda}} \leftarrow [\![\mathsf{pos}]\!]_{2^{\lambda}} [\![r]\!]_{2^{\lambda}}$ \\
    
    \KwRet{$[\![\rho(z)]\!]_{2^{\lambda}}$ }
    \caption{Secure Protocol() $\pi_\rho$ for Computing the Activation Function $\rho$.\newline 
    \textbf{Constraints:} all values in $\mathbb{Z}_{2^\lambda}$ are representations of fixed point approximations of real numbers s.t. the lowest $a$ bits represent the fractional component, the next $b$ bits represent the integer component and $\lambda > a + b$. Further, a negative value $x$ is represented as $2^{\lambda} - |x|$.
    \label{alg:secureactivation}}
\end{procedure}

\subsection{Secure Logistic Regression Training}
We now present our secure LR training protocol that uses a combination of the previously mentioned building blocks.

Notice that in the full gradient descent technique described in Algorithm \ref{alg:gradientdescent}, the only operations that cannot be performed fully locally by the data processors, i.e. on their own local shares, are:
\begin{itemize}
    \item The computation of the inner product in line 7
    \item The activation function $\rho$ in line 7
    \item The multiplication of $t_d-o_d$ with $d_{d,i}$ in line 9
\end{itemize}

Our secure LR training protocol $\plr$ (described in Protocol \ref{alg:securetraining}) shows how the secure building blocks described before can be used to 
securely compute these operations. The inner product is securely computed using $\pip$ on line 5, and since this involves multiplication on numbers that are scaled to a fixed-point representation, we truncate the result using $\ptrunc$. The activation function is securely computed using $\pi_\rho$ on line 6. The multiplication of $t_d-o_d$ with $x_{d,i}$ is done using 
%batch-$\pmul$ 
secure multiplication with batching
on line 11. Since this also involves multiplication on numbers that are scaled, the result is truncated using $\ptrunc$ in line 14. A slight difference between the full gradient descent technique described in Algorithm \ref{alg:gradientdescent} and our protocol $\plr$, is that instead of updating $\Delta w_i$ after every evaluation of the activation function, we batch together all activation function evaluations before computing the $\Delta w_i$. Since the activation function requires a bit-decomposition of the input, we can now make use of the efficient batch bit-decomposition protocol batch-$\pdecompopt$ (see Appendix) within the activation function protocol $\pi_\rho$.

The LR training protocol $\plr$ UC-realizes the logistic regression training functionality $\flr$. The correctness is trivial and the security follows straightforwardly from the UC-security of the building blocks used in $\plr$.

\begin{functionality}{Functionality $\flr$}
\textbf{Input:} Upon receiving a message from Alice/Bob with her/his shares of $([\![\boldsymbol{x}_d]\!],[\![t_d]\!])$ for the set of training examples $D$, record the shares, ignore any subsequent messages from that party and inform the other party about the receipt.\\
 \\
\textbf{Output:} Upon receipt of the inputs from both parties, locally perform the same  computational steps as $\pi_\rho$ using the secret sharings. Let  $[\![\boldsymbol{w}]\!]$ be the resulting vector. Before the deliver of the output shares, a corrupt party can fix the shares that it will get, in which case the other shares are adjusted accordingly to still sum to $\boldsymbol{w}$. The output shares are delivered to the parties.
\end{functionality}

\begin{procedure}[h]
%\SetAlgoLined
\SetKwInOut{Input}{Input}
\SetKwInOut{Output}{Output}
\SetKwFor{For}{for}{do}{}
\SetKwFor{Until}{until}{do}{}
\SetKwFor{ForE}{for each}{do}{}

\Input{$([\![\boldsymbol{x}_d]\!],[\![t_d]\!])$ for a set of training examples $D$; learning rate $\eta$; number of iterations $n_{iter}$. All secret sharings in the description of this protocol are in $\mathbb{Z}_{2^\lambda}$ and thus we simplify the notation to $[\![\cdot]\!]$.}
\Output{$[\![\boldsymbol{w}]\!]$ for a vector 
of weights $w_i$ that minimize the sum of squared errors over the training data}
\For{ $i \leftarrow 0$ \KwTo $m$}{
    $[\![w_i]\!] \leftarrow 0$
}
\For{$0$ \KwTo $n_{iter}$}{
    \ForE{$(\boldsymbol{x}_d,t_d)$ in $D$}{
        $[\![z_d]\!] \leftarrow \ptrunc(\pip([\![\langle \boldsymbol{w}\rangle ]\!] , [\![\langle \boldsymbol{x}_d\rangle ]\!]))$\\
        $o_d \leftarrow \pi_\rho([\![z_d]\!])$
    }
    \For{$i \leftarrow 0$ \KwTo $m$}{
            $[\![\Delta w_i]\!] \leftarrow 0$
    }
    \ForE{$(\boldsymbol{x}_d,t_d)$ in $D$}{
        $[\![\boldsymbol{v}_{\mathsf{diff}}]\!]\leftarrow \langle ([\![t_d]\!]-[\![o_d]\!]),...,([\![t_d]\!]-[\![o_d]\!])\rangle$ // vector of length $m$\\
        $[\![\boldsymbol{v}_{\mathsf{gradient}}]\!] \leftarrow $batch-$\pmul([\![\boldsymbol{v}_{\mathsf{diff}}]\!],[\![\boldsymbol{x}_d]\!])$\\
        $[\![\Delta \boldsymbol{w}]\!] \leftarrow [\![\Delta \boldsymbol{w}]\!] + [\![\boldsymbol{v}_{\mathsf{gradient}}\!]$
    }
    \For{$i \leftarrow 0$ \KwTo $m$}{
        $[\![\Delta w_i]\!] \leftarrow \ptrunc([\![\Delta w_i]\!])$\\
        $[\![w_i]\!] \leftarrow [\![w_i]\!] + \eta[\![\Delta w_i]\!]$
    }
}
\KwRet{$[\![\boldsymbol{w}]\!]$}
\caption{Secure Logistic Regression Training Protocol() $\plr$.\label{alg:securetraining}}
 
\end{procedure}

The following steps describe end-to-end how to securely train a LR classifier:

\begin{enumerate}
    \item The TI sends the correlated randomness needed for efficient secure multiplication to the data processors. Note that while our current implementation has the TI continuously sending the correlated randomness, it is possible for the TI to send all correlated randomness as the first step, and therefore can leave and not be involved during the rest of the protocol.
    \item Each data owner converts the values in the set of training examples $D$ that it holds to a fixed-point representation as described in Equation \ref{eq:fixedpoint}. Each value is then split into two shares, which are then sent to the data processor 1 and data processor 2  respectively.
    \item Each data processor receives the shares of data from the data owners. They now have secret sharings 
    $([\![\boldsymbol{x}_d]\!], [\![t_d]\!])$ of the set of training examples $D$. The learning rate $\eta$ and number of iterations $n_{iter}$ are predetermined and public to both data processors.
    \item The data processors collaborate to train the LR model. They both follow the secure LR training protocol $\plr$.
    \item At the end of the protocol, each data processor will hold shares of the model's weights $[\![w_i]\!]$. Each data processor sends their shares to all of the data owners, who can then combine the shares to learn the weights of the LR model. 
\end{enumerate}

\subsection{Cryptographic Engineering Optimizations}
\subsubsection{Sockets and Threading}
A single iteration of the LR protocol is highly parallelizable in three distinct segments: (1) computing the dot products between the current weights and the data set, (2) computing the activation of each dot product result, and (3) computing the gradient and updating the weights. In each of these phases, a large number of computations are required, but none have dependencies on others. We take advantage of this by completing each of these phases with thread pools that can be configured for the machine running the protocol. We implemented the proposed protocols in Rust; with Rust's ownership concept, it is possible to yield results from threads without message passing or reallocation. Hence, the code is constructed to transfer ownership of results at each phase back to the main thread to avoid as much inter-process communication as possible. Additionally, all threads complete socket communications by computing all intermediate results directly in the socket buffer by implementing the buffer as a union of byte array and unsigned 64-bit integer array. This buffer is allocated on the stack by each thread which circumvents the need for a shared memory block while also avoiding slower heap memory. The implementation of this configuration reduced running times significantly based on our trials. 

Further, all modular arithmetic operations are handled implicitly with the Rust API's Wrapping struct which tells the ALU to ignore integer overflow. As long as the size of the ring over which the MPC protocols are performed is selected to align with a provided primitive bit width (i.e. 8, 16, 32, 64, 128) it is possible to omit computing the remainder of arithmetic with this construction.

\section{Results}

\begin{table*}[h!]
    \caption{Accuracy and training runtime for LR models}
    \centering
    \begin{tabular}{lrrrrrr}
    \hline
    & \# features & \# pos. & \# neg. & \# of & 5-fold CV  & avg. \\
        &   & samples & samples & of iterations & accuracy & runtime\\
    \hline
    BC-TCGA & 17,814 & 422 & 48 & 10 & 99.58\% & 2.52 sec\\
    GSE2034 & 12,634 & 142 & 83 & 223 & 64.82\% & 26.90 sec\\
    \hline
    \end{tabular}
    \label{tab:results}
\end{table*}

\begin{table*}[h!]
    \caption{Runtime comparisons between SecureML and our work}
    \centering
    \begin{tabular}{lrrrrrr}
    \hline
    & BC-TCGA training & GSE2034 training & activation function \\
        & (online) & (online) & (one evaluation) \\
    \hline
    Our work & 2.52 sec & 26.90 sec & 0.030 ms\\
    SecureML & 12.73 sec & 49.95 sec & 0.057 ms \\
    \hline
    \end{tabular}
    \label{tab:compare}
\end{table*}

\begin{table*}[h!]
    \caption{Activation function runtimes}
    \centering
    \begin{tabular}{lrrr}
    \hline

    \# evaluations & avg. runtime & runtime per activation \\
    & & (runtime\slash \#eval)\\
    \hline
    $256$ & 9 ms & 0.035 ms \\
    $512$ & 16 ms & 0.031 ms\\
    $1024$ & 30 ms & 0.029 ms\\
    $2048$ & 59 ms & 0.028 ms\\
    \hline
    \end{tabular}
    \label{tab:activation}
\end{table*}

We implemented the protocols from the \textbf{Methods} section in Rust\footnote{https://bitbucket.org/uwtppml/idash2019}
and experimentally evaluated them on the BC-TCGA and GSE2034 data sets of the iDASH 2019 competition. Both data sets contain gene expression data from breast cancer patients which are normal tissue/non-recurrence samples (negative) or breast cancer tissue/recurrence tumor samples (positive) \cite{xie2016comparison}. We trained LR models on both data sets with a learning rate $\eta = 0.001$. We use a fixed number of iterations for each data set: 10 iterations for the BC-TCGA data set and 223 iterations for the GSE2034 data set. The accuracy of the resulting models, evaluated with 5-fold cross-validation, is presented in Table \ref{tab:results}, along with the average runtime for training those models. It is important to note that these are the same accuracies that are obtained when training in the clear, i.e.~there is \textit{no accuracy loss} in the secure version.

We used integer precision $b = 15$, fractional precision $a = 12$ and ring size $\lambda=64$ (these choices were made based on experiments in the clear as mentioned in the previous section).
We ran the experiments on  AWS c5.9xlarge machines with 36 vCPUs, 72.0 GiB Memory. Each of the parties ran on separate machines (connected with a Gigabit Ethernet network), which means that the results in Table \ref{tab:results} cover communication time in addition to computation time.
The results show that our implementation allows to securely train models with state-of-the-art accuracy \cite{xie2016comparison} on the BC-TCGA and GSE2034 data sets within about 2.52 seconds and 26.90 seconds respectively. 

A previous version of this implementation was submitted to the iDASH 2019 Track 4 competition. 9 of the 67 teams who entered Track 4 completed the challenge. Our solution was one of the 3 solutions who tied for the first place. Our implementation trained on all of the features for both data sets (no feature engineering is done), and generated a model that gave the highest accuracy, with runtimes that were well within the competition's limit of 24 hours. The implementation presented in the current work is further optimized in relation to the iDASH version and achieves far better runtimes.

We note that while SecureML differs from our work in their setup and cryptographic primitives, it shares many similarities to ours and reports a fast runtime such that we find it valuable as a standard to compare to. While SecureML does not originally use a TI to predistribute the multiplication triples, it would be easy to adapt their result to use a TI for that purpose. Therefore, in order to have a fair comparison, we compare our protocol runtime against only their online runtime (thus excluding their offline runtime). We evaluated our implementation's runtime against SecureML's implementation by running their implementation on the same AWS machines using the same data sets (see Table \ref{tab:compare} for runtime comparisons). For both data sets, our online phase runs faster than SecureML's online phase which trains BC-TCGA in 12.73 seconds and GSE2034 in 49.95 seconds.

We then compare online microbenchmark computation times. For the computation of the activation function, our run of the SecureML code reported around 0.057 ms to 0.059 ms for 1 activation, while our implementation completes 1024 evaluations in around 30 ms (0.029 ms per activation function). This makes our secure activation function implementation nearly twice as fast as SecureML's. Additionally, it eliminates the overhead of switching between Yao gates and additive secret sharing. 
Furthermore, our activation function runs more efficiently (per evaluation) the more evaluations of it need to be computed, due to the design of the batch bit-decomposition protocol. This is illustrated in Table \ref{tab:activation} where the calculated runtime per evaluation (runtime divided by number of evaluations) decreases as the number of evaluations increase.

%W (Not true anymore) e note that in SecureML, the runtimes for dot products and multiplications is optimized through vectorization (i.e. performing matrix-vector multiplications instead of multiple inner products). They report this vectorization improved the online time by around 2$\times$. They also reuse the random multiplication triples that are used to mask the data, which further decreases their computation and communication time. While our implementation does not take advantage of this vectorization or reuse, it is something that can be added in a future version to further improve our runtimes. \magenta{not true anymore: revise}

\section{Discussion}
Our runtime experiments on securely training a LR model show that it is feasible to train on data that includes a large number of attributes, as is common with genomic data. Given the high dimensionality of the genomic data, an interesting direction for future work would be the design of MPC protocols for privacy-preserving feature reduction. If any kind of feature reduction is used, it would result in a decrease in secure training runtime with a possibility for a slight decrease in the accuracy. We demonstrate this by choosing (in the clear) 54 features of the BC-TCGA data set that were part of the 76-gene signature described in \cite{wang2005gene}. Training on these 54 features, we get a 5-fold cross-validation accuracy of $98.93\%$ (training on all features produced $99.58\%$), and the average secure training time (of three runs) is 0.51 seconds, which is about a 2 second decrease from training on all 17,814 features. The genes in the GSE2034 data set are not labeled in a way where we can map them to the 76-gene signature to test the accuracy for a reduced number of features, but we test the runtime of training on 76 attributes and we get an average of 6.71 seconds, which is about a 20 second decrease from training on all 12,634 features. This shows that if feature reduction can be performed, runtimes can be improved while still being able to produce an accurate trained model.

Our main contribution is the proposal of the fastest implementation and protocol for privacy-preserving training of LR models. Our novelty points are the new protocol for privately evaluating the activation function $\rho$ which can be computed using only additive shares and MPC protocols, without using a protocol for secure comparison. We use $\rho$ as an approximation of the sigmoid function $\sigma$ since that is what is traditionally used in LR training, but $\sigma$ is also used as an activation function in neural networks. Therefore, our fast secure protocol for computing $\rho$ can also result in faster neural network training. While training neural networks are out of the scope of this paper, we note that our results can be applicable to those types of ML models as well.

\section{Conclusions}
In this paper, we have described a novel protocol for implementing secure training of LR over distributed parties using MPC. Our protocol and implementation present several novel points and optimizations compared to existing work, including: (i) a novel protocol for computing the activation function that avoids the use of full-fledged secure comparison protocols; (ii) a series of cryptographic engineering optimizations to improve the performance.

With our implementation, we can train on the BC-TCGA data set with 17,814 features and 375 samples with 10 iterations in 2.52 seconds, and we can train on the GSE2034 data set with 12,634 features and 179 samples with 223 iterations in 26.90 seconds.
A less optimized version of this implementation won first place at the iDASH 2019 Track 4 competition when considering accuracy and efficiency. Our solution is particularly efficient for LANs where we can perform 1024 secure computations of the activation function in about 30 ms. To the best of our knowledge, ours is the fastest protocol for privately training logistic regression models over local area networks.

\section*{Appendix}

\subsection*{Optimization of \pdecomp}

\subsubsection*{Overview and Previous Work}
The functionality $\fdecomp$ (described in Section \textbf{Methods}) is easily realized as an adder circuit that takes as inputs each bit of the additive shares of a secret sharing   $[\![x]\!]_{2^\lambda}$ in a large ring $\mathbb{Z}_{2^\lambda}$ and outputs an ``XOR-sharing" of the secret $[\![x_1]\!]_2, \ldots, [\![x_{\lambda}]\!]_2$. First, each party regards its share of $[\![x]\!]_{2^\lambda}$, denoted $x_i$, as an XOR-shared secret $[\![x_{i,1}]\!]_2, \ldots, [\![x_{i,\lambda}]\!]_2$ and passes it to the adder circuit. The adder circuit then computes the \emph{carry vector} which accounts for the rollover of binary addition. Adding this vector to all bitwise shares $[\![x_{1,j}]\!]_2, [\![x_{2,j}]\!]_2$ resolves the difference between $[\![x_{1,1}]\!]...[\![x_{1,\lambda}]\!] \oplus [\![x_{2,1}]\!]...[\![x_{2,\lambda}]\!]$ and the bit-decomposed secret $x$.

%When the carry vector combined with each bitwise sum $[\![x_{1,j}]\!]_2 \oplus [\![x_{1,j}]\!]_2$, the result $[\![x_1]\!]_2, \ldots,[\![x_{\lambda}]\!]_2$; the bit-decomposition of $[\![x]\!]_{2^\lambda}$.  

Naively, this carry vector can be obtained with linear communication complexity by means of ripple carry addition, as is described in Protocol \ref{prot:decomp}. But, it is possible to achieve logarithmic communication complexity and even constant complexity \cite{Toft2009ConstantRoundsAB} (though with worse performance than the logarithmic version for all reasonable bit lengths).

The highest performing realization of $\fdecomp$ for realistic bit lengths is based on a speculative adder circuit \cite{IEEETDSC:CDHK+17} in which at each layer the next set of carry bits are computed twice; once for each case that the previous carry bit had been 0 and 1. This protocol has $\lceil log(\lambda)\rceil + 2$ rounds of communication and requires a total data transfer of $4\lambda \lceil log(\lambda) \rceil + 6\lambda$ bits.

We propose a new, highly optimised protocol based on a matrix composition network that reduces the number of communication rounds by 1 (or 2, in special cases) and requires a small fraction of the aforementioned data transfer cost.

\subsubsection*{Matrix composition network}

To sum the binary numbers $a$ and $b$, the $i$-th bit is given by $s_i= a_i \oplus b_i \oplus c_{i-1}$, where $c_i = a_ib_i \oplus a_ic_{i-1} \oplus b_ic_{i-1}.$ In an alternate view, the carry can be seen to depend on two signals which in turn depend on $a$ and $b$. $\mathsf{Generate}$ ($g_i = a_ib_i$) creates a new carry bit at the $i$-th position, and $\mathsf{Propogate}$ ($p_i = a_i \oplus b_i$) perpetuates the previous carry bit, if it exists. In this representation, $s_i = p_i \oplus c_{i-1}$ and $c_i = g_i + p_ic_{i-1}$. This sum-of-products form of the expression for $c_i$ lends itself to a matrix representation

\[
\begin{bmatrix}
c_i \\ 1
\end{bmatrix}
=
\begin{bmatrix}
p_i & g_i \\
0 & 1 \\
\end{bmatrix}
\begin{bmatrix}
c_{i-1} \\ 1
\end{bmatrix}
=
M_i
\begin{bmatrix}
c_{i-1} \\ 1
\end{bmatrix}.
\]  

When matrices in the form of $M_i$ are composed, the lower entries remain unchanged. This implies that

\[
\begin{bmatrix}
c_i \\ 1
\end{bmatrix}
=
\begin{bmatrix}
p_i & g_i \\
0 & 1 \\
\end{bmatrix}
\begin{bmatrix}
p_{i-1} & g_{i-1} \\
0 & 1 \\
\end{bmatrix}
\begin{bmatrix}
c_{i-2} \\ 1
\end{bmatrix}
=
M_iM_{i-1}
\begin{bmatrix}
c_{i-2} \\ 1
\end{bmatrix}.
\]

Therefore, to compute all $c_i$, it is sufficient to compute the set of all matrix compositions 

\[ \left\{ \prod_{j=1}^{i} M_{j} \; \bigg\rvert \; 1 \leq i < \lambda \right\}.\]

Note that it is not necessary to compute the $\lambda$-th carry bit because $s_{\lambda}$ depends on $c_{\lambda-1}$. Treating the carry-in to the 1st bit as the vector $(0, 1)$, all $c_i$ can be derived implicitly from the upper right-hand entry of $M_{1.i}$ (here, $M_{1,i}$ denotes the matrix composed of all matrices $M_1$ through $M_i$, consecutively).

From the MPC perspective, this matrix composition requires two $\mathbb{Z}_2$ multiplications: $p_{i+1}p_i$ and $p_{i+1}g_i$ as seen in the equation below. The OR operation (+), which usually requires multiplication in MPC, is reduced to XOR based on the observation that $p_{i+1}$ and $g_{i+1}$ cannot both be true for a given $i$.

\[
\begin{bmatrix}
p_{i+1} & g_{i+1} \\
0 & 1 \\
\end{bmatrix} 
\begin{bmatrix}
p_i & g_i \\
0 & 1 \\
\end{bmatrix}
=
\begin{bmatrix}
p_{i+1}p_i & p_{i+1}g_i + g_{i+1} \\
0 & 1 \\
\end{bmatrix}
\]

The entire set of matrix compositions can be realized in a logarithmic depth network by, at the $i$-th layer, computing all compositions $M_{1.j}$ that require fewer than $2^{i-1}$ compositions. To set up conditions to allow us to minimize the total data transfer, the constraint is added that each $M_{1.j}$ should be the composition of the ``largest" matrix from the previous layer, $M_{1.2^{i-2}}$, with the remainder $M_{2^{i-2}+1.j}$. If $M_{2^{i-2}+1.j}$ doesn't exist in the network, it is added recursively following the same set of constraints. 

\begin{figure*}[h!]
    \centering
    \includegraphics[width=0.95\textwidth]{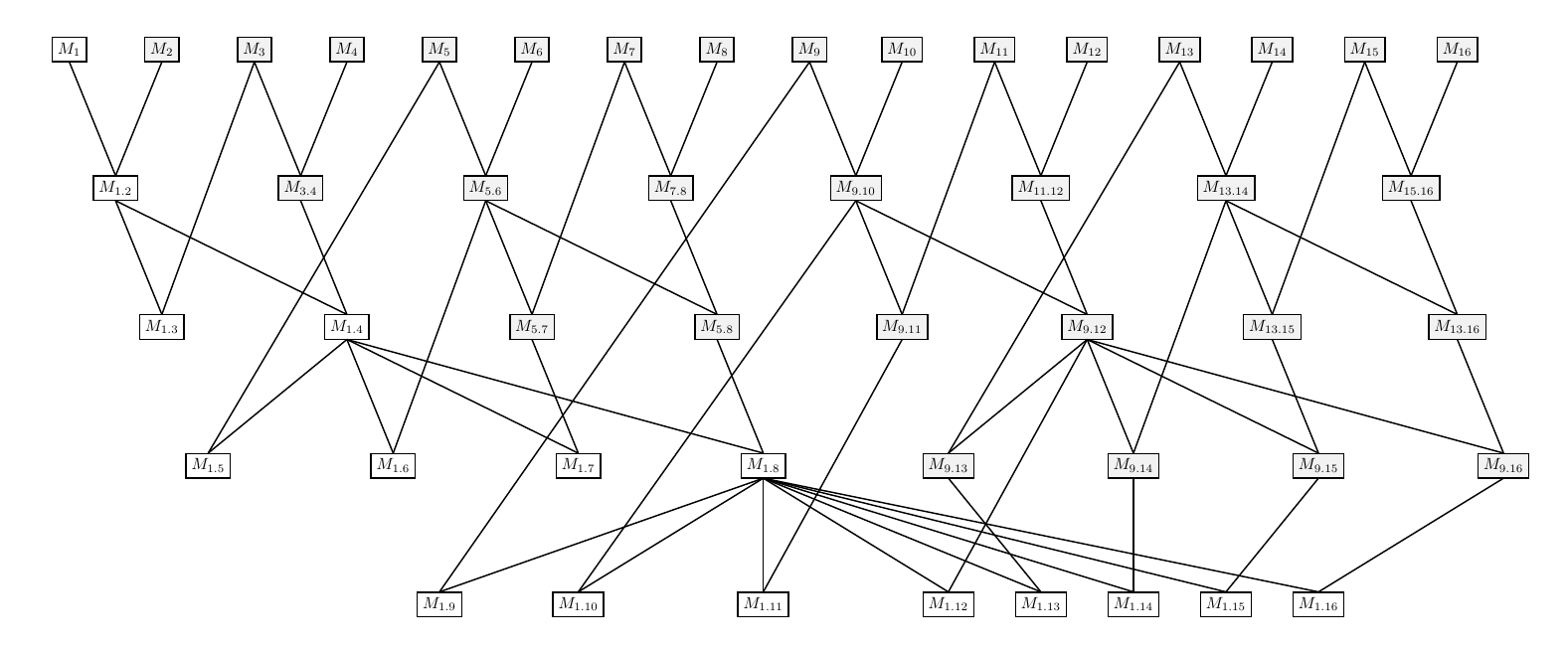}
    \caption{$\textsf{ComposeNet}_{\lambda}$ for $\lambda=17$. Computes the set of all matrix compositions $M_1, M_{1.2}, M_{1.3},\ldots,M_{1.(\lambda-1)}$. The notation $M_{i.j}$ means "the composition of all matrices $i$ through $j$." The greyed nodes are only used for intermediate computations and the white nodes are part of the solution set.}
    \label{fig:matcomp}
\end{figure*}

\figref{fig:matcomp} shows an example with $\lambda = 17$. This network is hereafter referred to as $\mathsf{ComposeNet}_p$ where $p$ is the highest order bit to decompose. The protocol description that follows considers only the case where $p = \lambda$, though the protocol functions the same for any $p \leq \lambda$. For instance, in Protocol \ref{prot:decomp}, when using $\pdecomp$ to find the $\mathsf{MSB}$ of a secret, it is sufficient to set $p = a + b + 1$. 

\begin{procedure}[h]
    \SetKwInOut{Input}{Input}
    \SetKwInOut{Output}{Output}

    \Input{ $[\![x]\!]_{2^{\lambda}}  $}
    \Output{$[\![x_1]\!]_2...[\![x_{\lambda}]\!]_2$}
    
    Party $i$ regards its share $x_i$ as $p_{i,1}, \ldots,p_{i,\lambda}$ s.t. \\
    $[\![ p_j ]\!]_2, = p_{1,j} \oplus p_{2,j}$ for $j=1,\ldots, \lambda$ \\
    Party 1 creates the sharing $[\![ g_{1,j} ]\!]_2=(p_{1,j}, 0)$.\\
    Party 2 creates the sharing $[\![ g_{2,j} ]\!]_2=(0,p_{2,j})$.\\
    $[\![ g_j ]\!]_2 \leftarrow [\![ g_{1,j} ]\!]_2[\![ g_{2,j} ]\!]_2$\\
    $[\![M_j]\!]_2 \leftarrow 
        \begin{bmatrix}
            [\![p_j]\!]_2 & [\![g_j]\!]_2 \\
            0 & 1 \\
        \end{bmatrix}$ for all $j$\\
    $\{ [\![M_{1.j}]\!]_2 | 1\leq j < \lambda\} \leftarrow 
    \mathsf{ComposeNet}_{\lambda}( [\![M]\!]_2 )$ \\
    $[\![c_j]\!]_2 \leftarrow$ the upper right entry of $[\![M_{1.j}]\!]_2$ \\
    $[\![s_1]\!]_2 \leftarrow [\![p_1]\!]_2$  \\
    $[\![s_j]\!]_2 \leftarrow [\![p_j]\!]_2 \oplus [\![c_{j-1}]\!]_2$ for all $j > 1$ \\

    \KwRet{$[\![s_1]\!]_2...[\![s_{\lambda}]\!]_2$}
    \caption{Secure Protocol() $\pdecompopt$ for computing \fdecomp more efficiently.} 
\end{procedure}

\subsubsection*{Efficiency discussion}

The setup phase prior to the call to $\mathsf{ComposeNet}_{\lambda}$ requires $\lambda$ multiplications over $\mathbb{Z}_2$ to compute all $[\![g_j]\!]$. This corresponds to one communication round and $2\lambda$ bits of data transfer. 

A call to $\mathsf{ComposeNet}_{\lambda}$ has communication complexity corresponding to the depth of the network, $\lceil \log(\lambda-1)\rceil$, and $\frac{\lambda}{2}$ multiplications over $\mathbb{Z}_2$ per layer, with fewer on the final layer when $\lambda - 1$ is not a power of 2. However, due to the fact that the matrices at each node of $\mathsf{ComposeNet}_{\lambda}$ are reused extensively and known to not change value, the Beaver Triples used to mask the matrices can be desgined to contain redundancies to minimise the data transfer at each layer \cite{mohassel2017secureml}. By re-using correlated randomness where information leakage is not possible, only $\frac{\lambda}{2}-(2^{i-i}-1)$ masks need to be transferred at depth $i$, for $i > 0$. At depth 0, there are $\lambda$ masks; one for each matrix. Each matrix mask is 2 bits (one for each of the $\mathsf{Propogate}$ and $\mathsf{Generate}$ bits), so the total data transfer is $2\lambda + 2\sum_{i=1}^{\lceil log(\lambda-1)\rceil-1} (\frac{\lambda}{2}+1-2^{i-1})$.

The recombination phase after $\mathsf{ComposeNet}_{\lambda}$ is computed has only local computations and thus contributes nothing to the complexity.

Combining all phases, we see that $\pdecompopt$ has a communication cost of $\lceil \log(\lambda-1)\rceil + 1$ and a total data transfer cost of $4\lambda + 2\sum_{i=1}^{\lceil \log(\lambda-1)\rceil-1} (\frac{\lambda}{2}+1-2^{i-1})$ bits. Comparing with the speculative adder's performance, the number of communication rounds is decreased by 1 in all cases and 2 in the case that $\lambda-1$ is a power of 2. The total data transfer cost
%in expanded form
%3\lambda + (\lambda+2)\lceil log(\lambda-1)\rceil - 2^{ \lceil log(\lambda-a)\rceil } - 1$ 
has roughly $\frac{1}{3}$ the data transfer rate of the previous work at $\lambda=8,16$. For higher all bit lengths, the ratio quickly converges near $\frac{1}{4}$.  

\subsubsection*{Implementation and Batching}
$\mathsf{ComposeNet}_{\lambda}$ can be implemented efficiently as a set of index pairs that correspond to the positions of the $\mathsf{Propogate}$ and $\mathsf{Generate}$ bits that need to be combined at each layer. Once per layer, all products $p_{i+1}p_i$, $p_{i+1}g_i$ can be computed in a single call to \pmul by taking the bitwise product between the concatenations $p_{i+1}||p_{i+1}$, $p_i||g_i$ and splitting the result.

Extending to the case that many values need to be bit decomposed at the same time (as in Protocol 6), a vector of inputs can be decomposed ``in parallel" by taking vertical slices over the $\mathsf{Generate}$ and $\mathsf{Propogate}$ bits of each element and re-packing them into a transposed form. In this way, each layer of $\mathsf{ComposeNet}_{\lambda}$ can operate on a vector of matrices (represented as two lists of bit slices) to produce a vector of matrix compositions.  This method has no effect on the number of rounds of communication and the total data transfer scales linearly with the length of the input vector.  
 
\section*{Acknowledgements}
The authors want to thank P. Mohassel for making the SecureML code available that was used for the experimental comparison in the Section \textbf{Results}.

\end{document}